\documentstyle[aps,prl,preprint]{revtex}
\title{\bf \Huge Strong to fragile transition in a model of liquid silica}

\author{Jean-Louis {\sc Barrat}\thanks{D\'epartement de Physique des Mat\'eriaux  
(UMR CNRS 5586), Universit\'e Claude Bernard, 43 bd. du 11 novembre 1918, F-69622 
Villeurbanne Cedex, FRANCE} 
James {\sc Badro}\thanks{Laboratoire de Sciences de la 
Terre (URA 726), Ecole Normale Sup\'erieure de Lyon, 46 All\'ee d'Italie 
F-69364 Lyon Cedex 07, FRANCE}, 
and Philippe {\sc Gillet}\thanks{Laboratoire de Sciences de la 
Terre (URA 726), Institut Universitaire de France, Ecole 
Normale Sup\'erieure de Lyon, 46 All\'ee d'Italie,
F-69364 Lyon Cedex 07, FRANCE}}

\begin{document}

\maketitle

\begin{abstract}
The transport properties of an ionic model for liquid silica \cite{vanbeest}
at high temperatures and pressure are investigated using molecular dynamics 
simulations.  With increasing pressure, a clear change from
"strong"  to "fragile"  behaviour (according to Angell's classification
of glass-forming liquids) is observed, albeit only on the small viscosity
range that can be explored in MD simulations.   This change is related to
structural changes, from an almost perfect four-fold coordination
to an imperfect five or six-fold coordination.  

\end{abstract}

\newpage

More than 10  years ago, Angell \cite{angell} proposed a 
classification of glass forming liquids according to their "fragility",
a  concept that has since then proved extremely useful  for our understanding
of these systems. Broadly speaking, "strong" glass formers are characterized by weak
discontinuities of their thermodynamic properties at the glass transition, 
and an almost Arrhenius behaviour of their transport properties (viscosity)
as a function of temperature.  The corresponding activation energy $E_a$
can be assigned to a typical microscopic "event' that controls transport
properties, such as breaking of a  "bond". The archetype of "strong liquids"
is $SiO_2$, with a perfectly Arrhenius behaviour of the viscosity
in the range $1500 < T< 2500$. The  activation energy is in that case
about 70000K, and is supposed to correspond to the energy
for breaking an $Si-O$ bond.

Fragile liquids, on the other hand, display a 
strongly non Arrhenius behaviour of their transport properties. 
The flow in these systems can not be traced back to
a single microscopic event, but is of a much more collective nature.
Typical fragile liquids are organic or ionic systems with a much weaker
local organization than in "strong" liquids.
it is now widely accepted that at least for relatively low viscosities,
the transport properties in these systems are well described by the so called
"mode-coupling" theories \cite{Gotze}.

It has long been suspected, and also shown
in some pioneering MD simulations \cite{cheeseman}, that the transport properties 
of $SiO_2$ could exhibit an unusual behaviour as a function of pressure.
in particular, a nonmonotonous, "waterlike" behaviour of the diffusion
coefficient is expected due to the tetrahedral network structure of 
molten $SiO_2$.  It is also expected \cite{barratklein,angell} that the disruption
of this network by an applied pressure will cause a change in behaviour from
"strong" to "fragile".

In this work, we present a systematic  Molecular Dynamics (MD)
study of the transport properties
in a {\it purely ionic } model of $SiO_2$ \cite{vanbeest},
which is an improvment of an earlier model by Tsuneyuki 
and coworkers
\cite{tsuneyuki}. Both models 
 have now been studied in much detail for their static properties
\cite{dellavalleandersen,kob,tseklug,Guillot,badro}. Transport properties at
zero pressure have also been briefly investigated in \cite{dellavalleandersen2}, 
and their pressure dependence was considered in \cite{yuenrustad}.

The system we consider is made up of 216  $SiO_2$ units (in some
cases this number was increased to 512 in order to check the size
dependance of the results). The equations of motion are integrated using 
the Verlet algorithm with temperature control \cite{AT87}. Four series of runs
at four different densities  (2.2, 3.1, 3.3 and 4.2 g/cc) were carried out. For 
each density, the sample is equilibrated at high temperature (5000K), then 
cooled stepwise at an average cooling rate of about $10^{12} K.s^{-1}$. At each new temperature,
the system is equilibrated during 0.25 to 0.5 nanosecond.  As usual \cite{barratklein}, this equilibration time limits the temperature range for which the system can be considered to
be at thermodynamic equilibrium. By analogy with the situation in 
real systems, the lowest temperature at which the system 
is equilibrated for an equilibration time of 0.5 nanosecond 
defines the "computer glass transition temperature" for our simulation.
 In terms of viscosities, this corresponds to viscosities close
to 1. Pa.s..

The  self diffusion coefficients of $Si$ and $O$, $D_{Si}$ and 
$D_O$, were computed from the mean squared displacement 
of the corresponding ions. The shear viscosity $\eta$ was computed from 
the stress-stress autocorrelation function using the usual 
Kubo formula \cite{barratklein}.  The results for $\eta$
and $1/D_{Si}$ are shown in figures 1 and 2,  in the form
of an Arrhenius plot.  As usual, the error bars on
$\eta$  (estimated from the contributions of the 3 components
of the stress tensor) are large,  typically 100\% for the larger viscosities. 
The results, however, are unambiguous. At the lowest density (2.2 g/cc), 
the Arrhenius plot is almost a  perfect straight line, 
with an activation energy 
about 40000 Kelvins. As previously noted by Della Valle and Andersen 
\cite{dellavalleandersen}, this activation energy is smaller than the 
experimentally observed one.  Most likely, this constitutes an inadequacy 
of the purely ionic model. Another interesting possibility, however, is 
 that the difference is due to the different temperature range investigated. 
Experiments are done at
low temperatures, while simulation investigates high temperatures.
In fact, a slight curvature of the Arrhenius plot is perceptible 
if the last point, corresponding to a temperature of 3200K, is included.
This would mean that the   activation "energy"   changes with temperature.
Unfortunately, the error bar on this point is large, due to the very small
diffusivity. 
High temperature experiments on $SiO_2$ would be useful
to clarify this issue.
As the density is increased,   a decrease in $\eta$ and $1/D_{Si}$ is observed,
while the Arrhenius plot displays a noticeable curvature. 
At  the higher density, the viscosity  has increased close to its low density
value, but the Arrhenius plot is  clearly bent, in a way characteristic of 
fragile liquids. The nonmonotonous variation of $D_{Si}$ with pressure is
illustrated in figure 3 for a fixed temperature ($T=4000K$). As pressure is
increased,  $D_{Si}$ increases by almost an order of magnitude,
with a maximum in the 10-15GPa range. This effect was already observed in
\cite{Lasaga} (with a different model) and \cite{yuenrustad}. The size effect reported in the latter  paper is not observable in
our results, as seen from the two points in figure 3 obtained with a 
1536 particles system.

Diffusion and vicosity are often related by  introducing a "Stokes Einstein diameter"
$d_{SE}$ of the atoms,  defined as $d_{SE}= k_BT / 3\pi \eta D$.
 This quantity, shown (for the $Si$ ions)  in figure 4 for the four densities
investigated, is seen to be fairly independent of temperature.  The scatter
observed at low temperature for the lower density is probably due to the
large error bars on the corresponding viscosities.
At the lowest density, $d_{SE}$ is large (3 to 4 \AA), but becomes much smaller, and
almost independent of density(1 to 1.5 \AA), for the higher densities. 
 This illustrates
the very peculiar character of low density $SiO_2$, with a perfectly tetrahedral
coordination.  The evolution of the local structure with pressure
is illustrated in figure 5, which shows the proportion of four, five and six fold 
coordinated $Si$ as a function of pressure at $T=4000K$.  In the standard 
picture of diffusion in $SiO_2$ \cite{cheeseman},
 five-fold coordinated $Si$ are often described as "defects" of the tetrahedral 
network
that favour diffusion.   Figure 5 shows that the proportion of these  "defects"
increases very rapidly with pressure, and that in the vicinity of the diffusion
maximum a large majority of $Si$ ions are indeed five-fold coordinated.
Hence it seems unlikely that a "defect based" picture of $SiO_2$ can 
be useful in describing the pressure dependance of its properties, except
perhaps at very low pressures.  Figure 5 also shows the evolution of 
coordination with pressure, from fourfold at $P=0$ to fivefold in the 
10-15 GPa range and sixfold above 20 GPa.  The low pressure four fold coordination,
however, is very special in the sense that it is alost perfect,
with more than 99\% of the ions.  At higher pressures, the coordination
is much less well defined.

Finally, the influence of the local structure on self diffusion is 
illustrated in figure 6.  The mean squared displacement of $Si$ ions
with different initial
coordinations (3, 4, or 5) is plotted as a function of time. For long times, 
these quantities become parallel straight lines. However at short times,
the difference between the different curves is  a measure of the influence
of coordination on the diffusion. This role is very clear in 
low density $SiO_2$,  where three  or five fold coordinated $Si$  (defects)
diffuse
initially much faster than the four-coordinated $Si$. At a density of 2.8g/cc,
(corresponding to about 6Gpa in figure 3 and 5), the role of initial coordination
has become almost negligeable.  The overshoot in the mean squared displacement that exists at low densities and disappears at higher densities
was noticed in \cite{Angellshao} and associated to the existence of 
a "boson peak". This overshoot, however, might be associated to a
finite size effect \cite{Kob_Angell}.

In summary, our results indicate that the disruption of the tetrahedral network
of low density $SiO_2$ by the application of
pressure very rapidly induces a change from strong to fragile behaviour. 
It must be remembered, however, that the change observed here corresponds
to a viscosity variation of less than three decades. Experimentally, the same 
kind of change could be expected to tkae place in a much larger viscosity domain,
covering more than ten decades. 
In the vicinity of the diffusivity maximum, 
 the behaviour is already that of a fragile liquid. 
The role of five or three fold coordinated "defects"   in the diffusion 
becomes negligeable before the diffusivity maximum is reached. 

These findings corroborate those obtained with other ionic
models of $SiO_2$ \cite{Angellshao}. They confirm the richness 
of these simple models.  Beyond the fact that they provide a 
relatively accurate description of silica, a great interest of these models
lies in the fact that they involve simple pair potentials, for which 
accurate theoretical methods have been devised.  They could  therefore provide 
a useful benchmark for theoretical studies of strong/fragile transitions.
Finally, the connection between these results and experiment
could be made in the high temperature-high pressure range 
(typically $T\sim 2500K, \ P\sim 20 GPa$. This range can be explored experimentally,
and corresponds (according to the model) to  diffusion constants and viscosities measurable
on the simulation time scale.  A comparison with experimental data would allow 
a better  evaluation of 
 the usefulness of such models in the prediction of dynamical properties.

{\bf Acknowledgments:} The calculations were made possible by 
computer time allocated by the Pole Scientifique de Mod\'elisation
Num\'erique at ENS-Lyon. Useful discussions with W. Kob are acknowledged.

\newpage

\centerline{\bf FIGURE CAPTIONS}

{\bf Figure 1:} Arrhenius plot of $1/D_{Si}$ at four different densities. Open dots: 
$\rho=2.2g/cc$.  Open stars: $\rho=3.3 g/cc$  Stars:  $\rho=3.7g/cc$.
 Squares: $\rho=4.2g/cc$. The lines are a guide to the eye.

{\bf Figure 2:} Same as figure 1, for the viscosity.
 The symbols are as in figure 1.

{\bf Figure 3:} Pressure dependance of the silicon diffuion constant
at $T=4000K$. The points corresponds to  densities
$\rho=2.2, 2.5, 2.8, 3.1,3.4,3.7,3.9,4.2 g/cc$. 
For $\rho=2.5g/cc$ and $\rho=3.4g/cc$ two different system sizes
(N=256  and N=512 $SiO_2$ units ) were considered.

{\bf Figure 4:} Stokes-Einstein diameter of the 
silicon ions as a function of temperature,
for the same four densities. as in figures 1 and 2.
Symbols as in figure 1 and 2.

{\bf Figure 5:}
Fraction of 4,5,and 6 coordinate silicon ions
as a function of pressure, for $T=4000K$. 
The densities are the same as in figure 3.. Inset: the equation
of state of the simulated system at T=4000K

{\bf Figure 6:} Mean squared displacement of silicon ions that are
initially 3,4,and 5 fold coordinated. Dots: fourfold coordinance. Long dashes:
fivefold coordinance. Short dashes: threefold coordinance. From top to
bottom: $\rho=2.2g/cc$, $\rho=2.8g/cc$, $\rho=3.1 g/cc$. $T=4000K$.


\newpage
\begin{thebibliography}{99}


\bibitem{vanbeest} B.W. van Beest, G.J. Kramer, R.A. van Santen,
"Force fields for silicas and aluminophosphates based on abinitio
   calculations."
Phys, Rev. Lett. {\bf 64}, 1955 (1990)


\bibitem{angell} C.A. Angell, "Strong and fragile liquids",
in "Relaxation in Complex Systems",
(K. Ngai, G.B. Wright eds), US Dpt of Commerce, Springfield (1985)


\bibitem{Gotze} W. G\"otze, L. Sj\"ogren,
 "Relaxation processes in supercooled liquids."
     Reports on progress in Physics,  {\bf 55}, 241 (1992)


\bibitem{cheeseman} C.A. Angell, P.A. Cheeseman, S. Tammadon, "Pressure enhancement of ion mobilities in liquid silicates from computer 
simulation studies"
Science, {\bf 218}, 885 (1982)

\bibitem{barratklein} J.L. Barrat, M.L. Klein,   "Molecular dynamics simulations of supercooled liquids near the glass
   transition."
Ann. Rev. Phys. Chem., {\bf 42}, 23  (1991)

\bibitem{tsuneyuki} S. Tsuneyuki, M. Tsukada, H.  Aoki, Y. Matsui
Phys. Rev. Lett. {\bf 61}, 869 (1988)

\bibitem{dellavalleandersen}  R.G. Dellavalle, H.C. Andersen
    "Test of a pairwise additive ionic potential model for silica."
J. Chem. Phys. {\bf 97},  2682 (1992)

\bibitem{kob} K. Vollmayr, W. Kob, K. Binder, "Cooling rate effects
in amorphous silica: a computer simulation study", 
preprint 1996


\bibitem{badro} J. Badro. J-L. Barrat, P. Gillet,
"Numerical simulation of alpha-quartz under nonhydrostatic compression -
   memory glass and five-coordinated crystalline phases."
 Phys. Rev.
Lett. {\bf 76}, 772 (1996)

\bibitem{tseklug}  J.S. Tse, D.D. Klug,
  "The structure and dynamics of silica polymorphs using a 2-body effective
    potential model."
J. Chem. Phys. {\bf 95}, 9176 (1991)

\bibitem{Guillot} B. Guillot, Y. Guisani, to be published

\bibitem{AT87} M. Allen, D. Tildesley, {\it Computer simulation
of fluids} (Oxford University Press, New-York, 1990)

\bibitem{Lasaga} J.D. Kubicki, A.C. Lasaga, 
     "Molecular dynamics simulations of $SiO_2$ melt and glass: ionic and
     covalent models."
Am. Mineral. {\bf 73}, 941 (1988)

\bibitem{dellavalleandersen2}  R.G. Dellavalle, H.C. Andersen
"Molecular dynamics simulation of silica liquid and glass."
J. Chem. Phys. {\bf 97},  2682 (1992)

\bibitem{yuenrustad} J.R. Rustad, D. Yuen, F.J. Spera,
     "Molecular dynamics of liquid $SiO_2$ under high pressure."
Phys. Rev. {\bf A42}, 2081 (1990)

\bibitem{Angellshao} C.A. Angell, P.H. Poole, J. Shao,
     "Glass-forming liquids, anomalous liquids, and polyamorphism in liquids and
   biopolymers." Il Nuovo Cimento {\bf 16D},
993 (1994).


\bibitem{Kob_Angell} J. Horbach, W. Kob, K. Binder, C.A. Angell,
"Finite-size effects in computer simulations of the dynamics
of strong glass formers", preprint 1996.

\end{thebibliography}
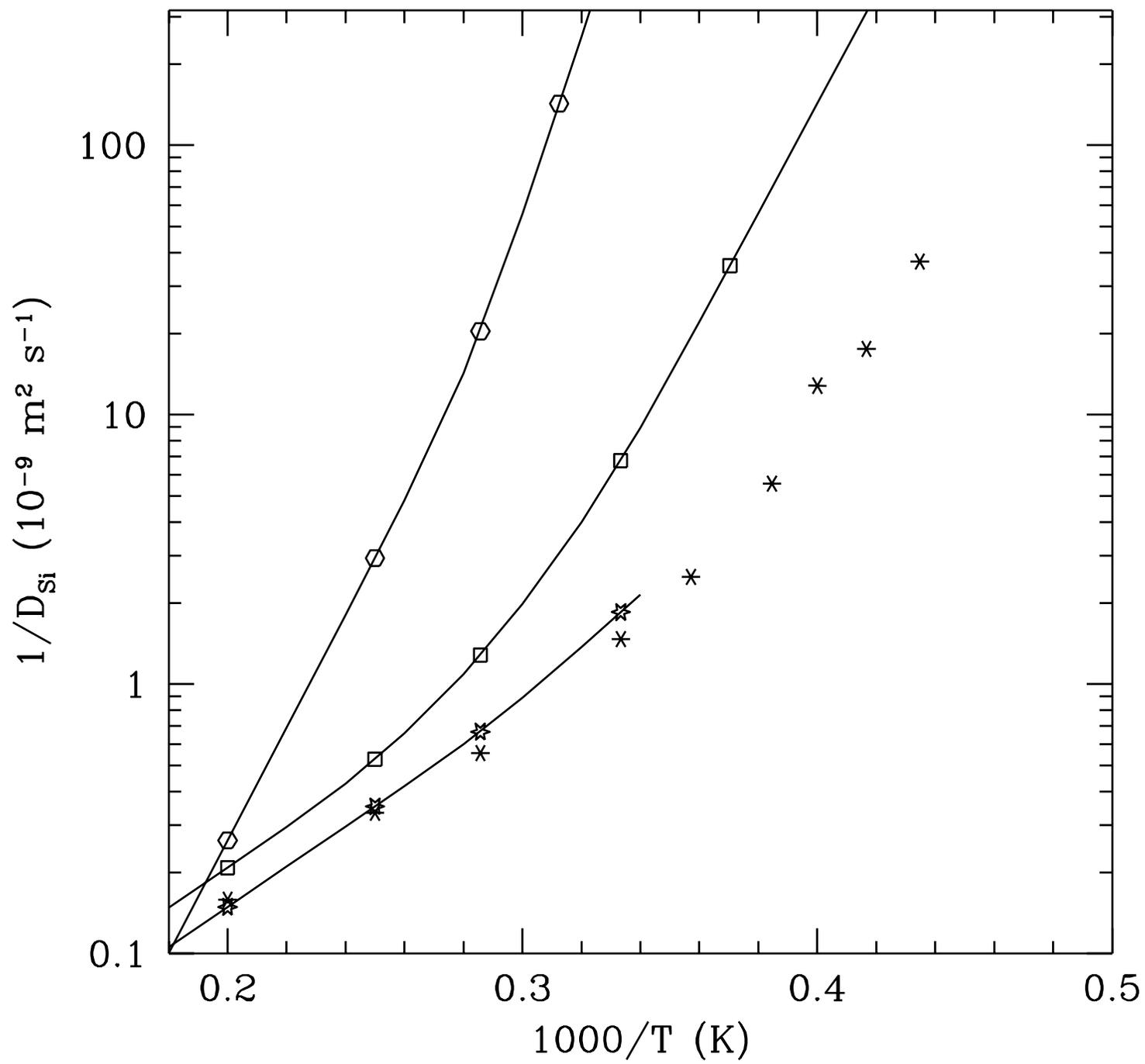
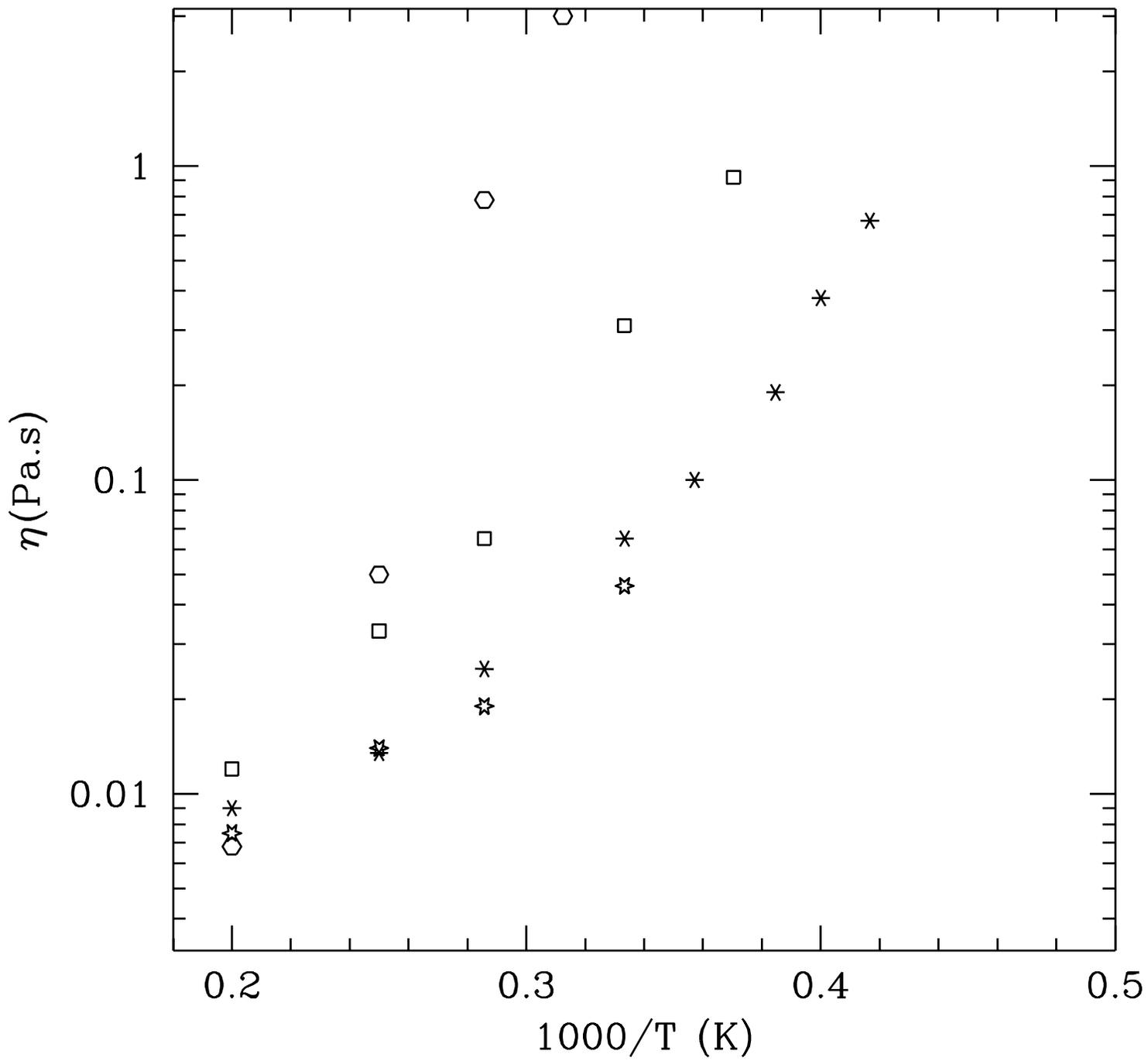
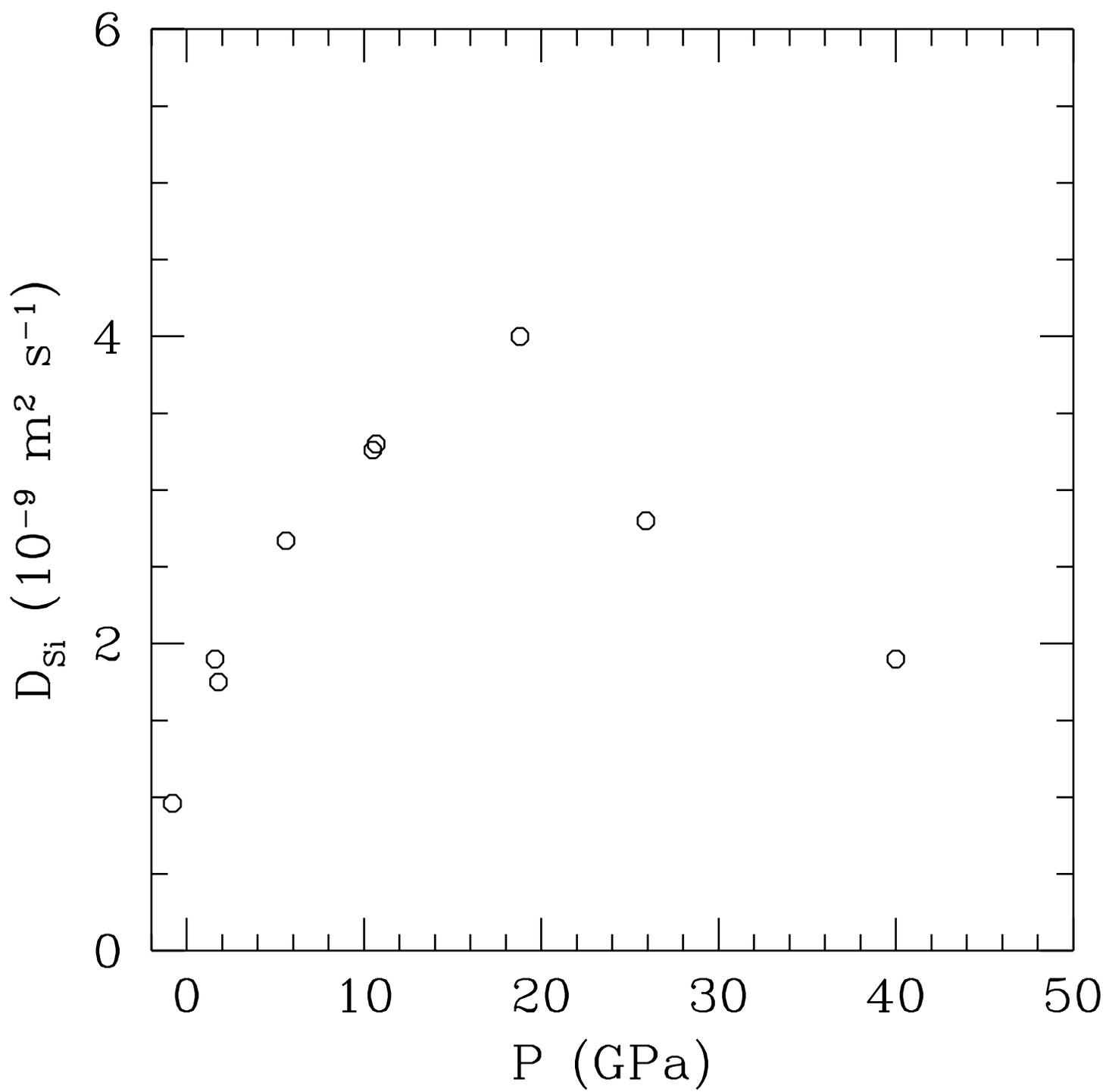
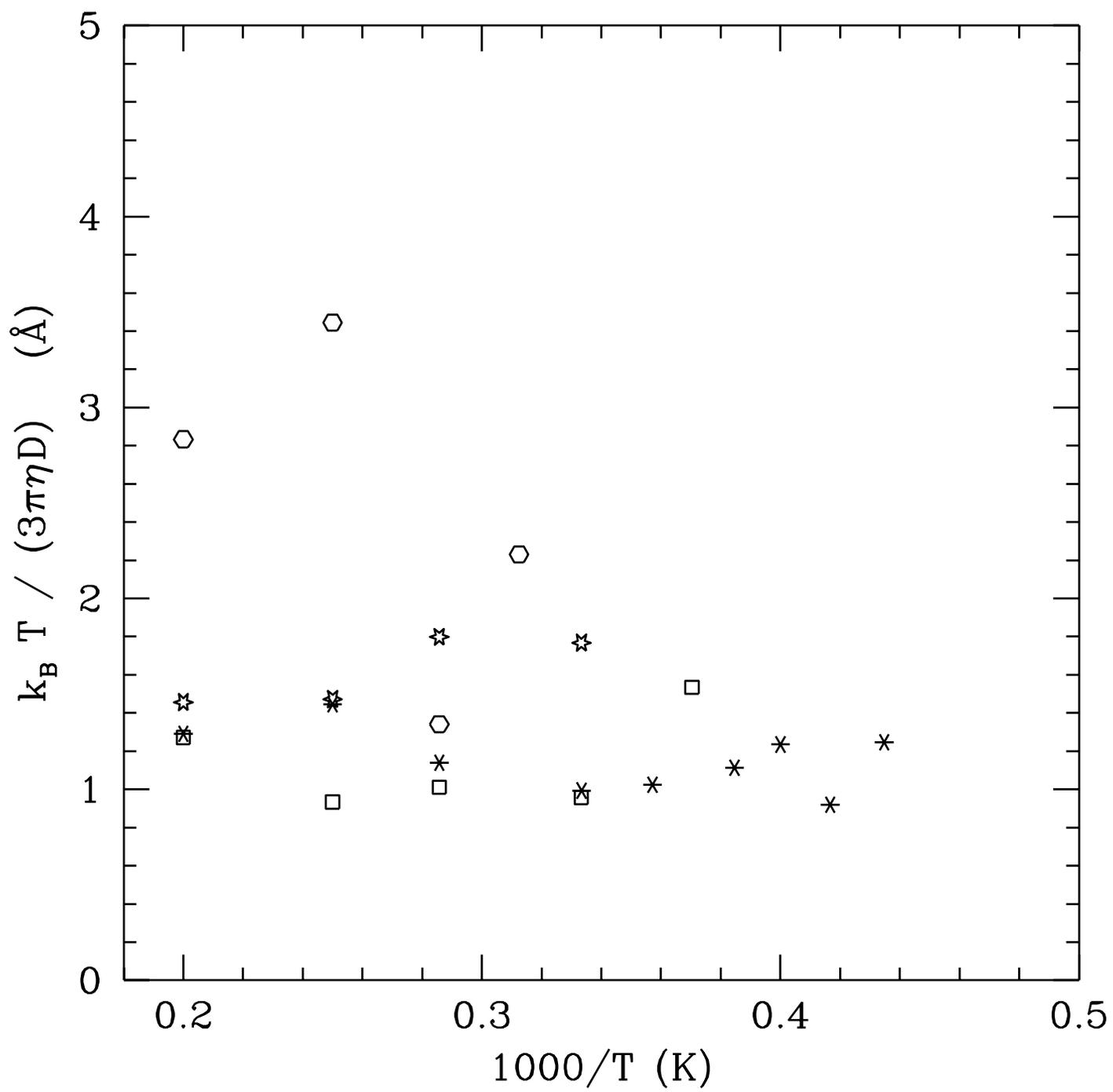
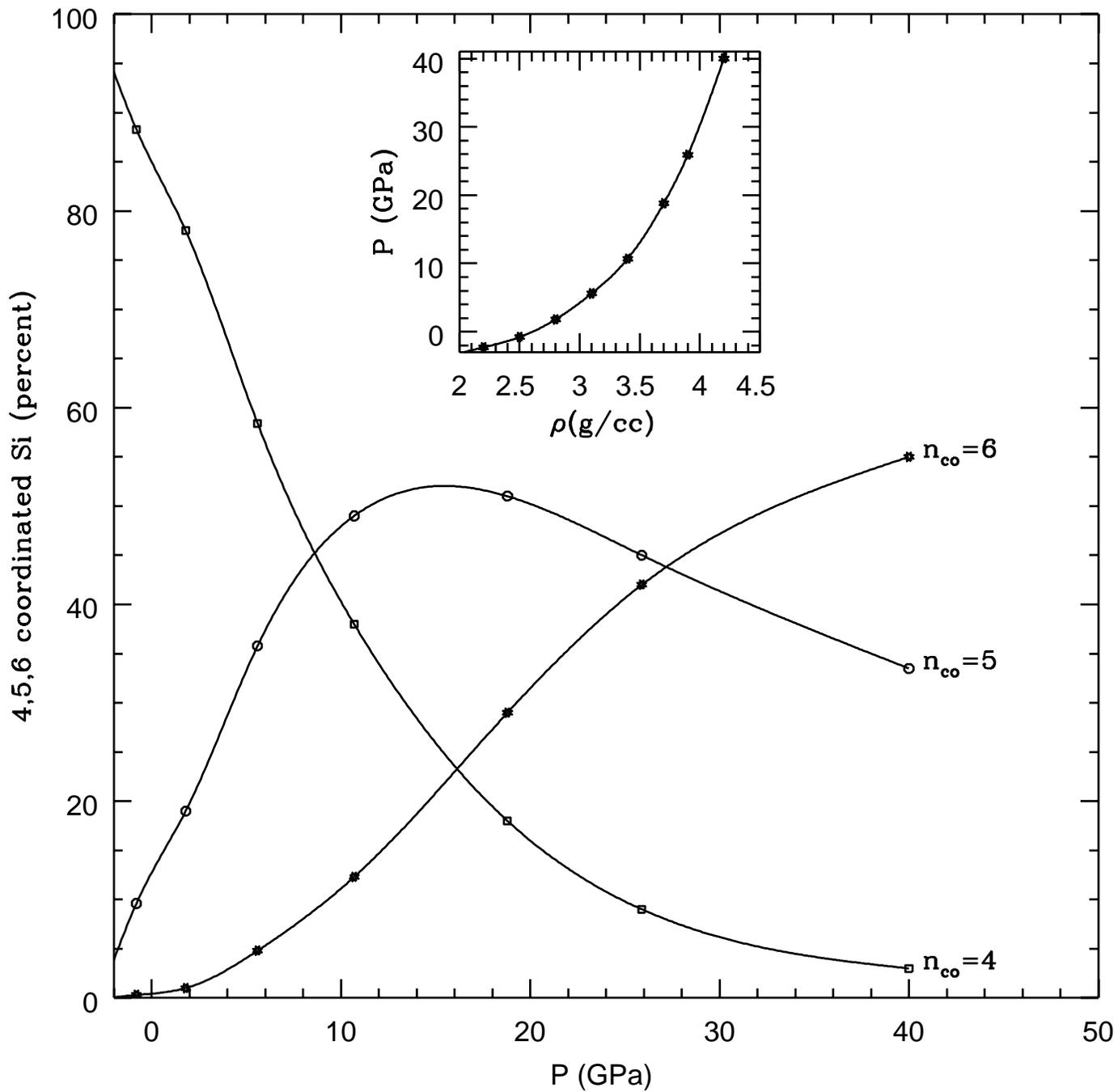
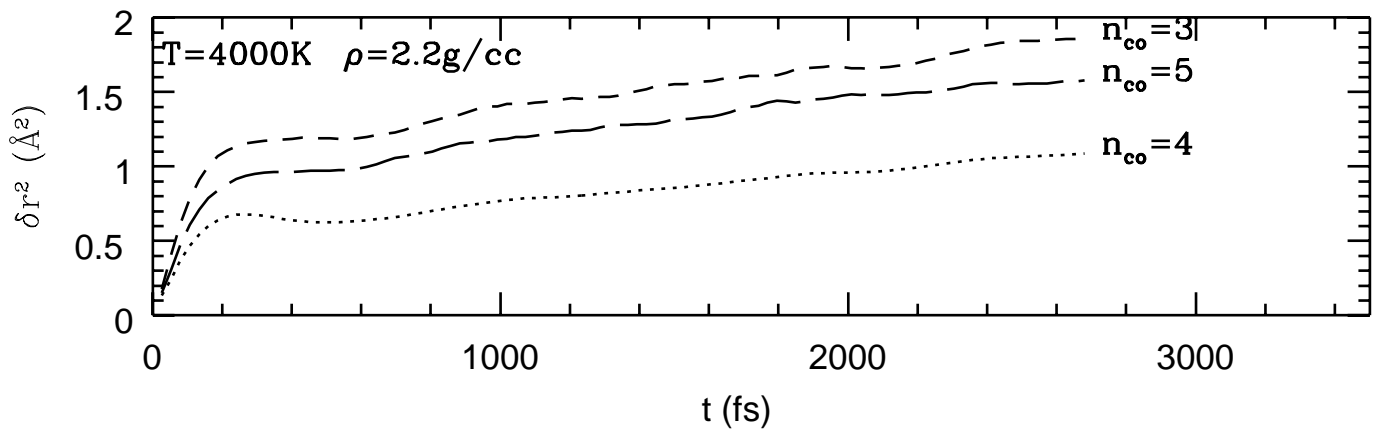
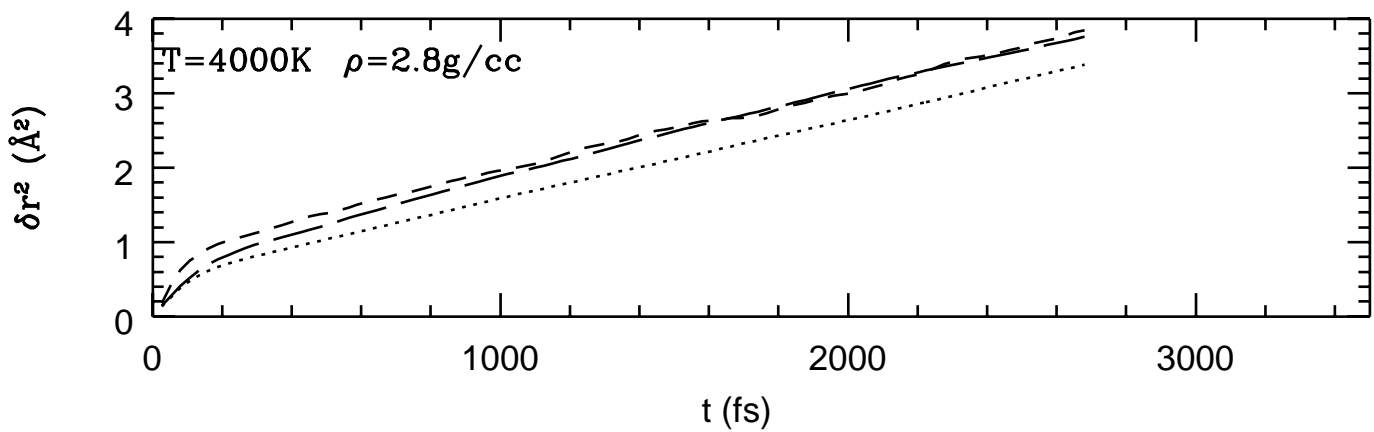
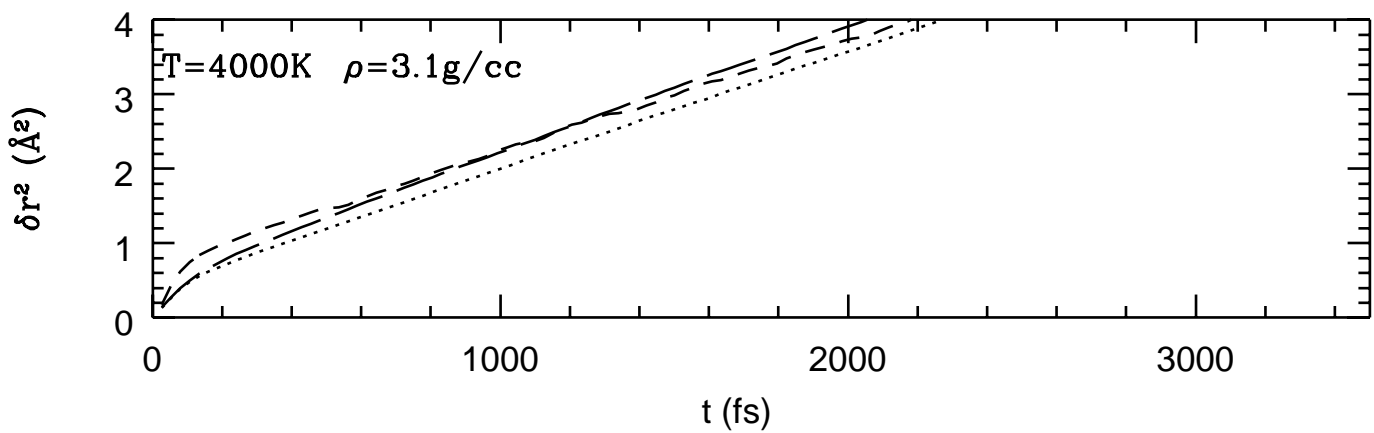
\end{document}